\documentclass[aps,prl,twocolumn,epsfig,amsfonts,amsmath,amstex,floatfix]{revtex4}

\usepackage{bm}
\usepackage{braket}
\usepackage{graphicx}
\usepackage{hyperref}

\newcommand{\Tr}{\operatorname{Tr}}

\newcommand{\cond}{\mathrm{cond}}
\newcommand{\ext}{\mathrm{ext}}
\newcommand{\inte}{\mathrm{int}}

\usepackage{bbold}
\usepackage{color}

\usepackage[]{amsmath}
\usepackage{braket}
\usepackage{amsfonts,amsmath}
\usepackage{epsfig,amsmath}
\usepackage{graphicx}% Include figure files
\usepackage{dcolumn}% Align table columns on decimal point
\usepackage{bm}% bold math %\bibliographystyle{apsrev}
\usepackage{bbold}
\newcommand{\beq}{\begin{equation}}
\newcommand{\eeq}{\end{equation}}
\newcommand{\beqa}{\begin{eqnarray}}
\newcommand{\eeqa}{\end{eqnarray}}

\newcommand{\ra}{\rangle}

\begin{document}

\title{Parameter-Decoupled Quantum Superresolution without Multiparameter Estimation}

\author{Xiao-Feng Qian}\email[]{xqian6@stevens.edu}
\affiliation{Department of Physics, and Center for Quantum Science and Engineering, Stevens Institute of Technology, Hoboken, NJ 07030, USA}

\date{\today}

\begin{abstract}
Quantum superresolution promises to resolve closely spaced sources beyond the diffraction limit, but existing approaches generally rely
on idealized source properties or require simultaneous estimation of multiple coupled parameters of realistic sources. Here we introduce a parameter-decoupled
measurement that determines the separation of realistic passive sources without estimating their unknown brightness imbalance, mutual coherence, or relative phase. A complete four-parameter quantum Fisher information analysis identifies the separation information that remains accessible in the presence of these nuisance parameters. We then construct a directly measurable log-probability invariant from  projection channels whose conditional statistics depend only on the separation. Broad classes of channel pairs satisfy the resulting decoupling condition. For a Gaussian point-spread function, the lowest-order  implementation asymptotically attains the four-parameter quantum limit per incident signal in the sub-Rayleigh regime. This approach converts a realistic multiparameter imaging problem into an operationally single-parameter measurement, providing a practical route toward quantum-enhanced resolution of passive spatial, temporal, and spectral signals.
\end{abstract}

\maketitle

\section{Introduction.}
Resolving closely spaced source signals is a canonical problem in optical imaging. Under direct intensity detection, the images of two nearby sources increasingly overlap and become indistinguishable as their separation decreases, giving rise to the Abbe-Rayleigh resolution limit \cite{abbe1873beitrage, rayleigh1879xxxi,born1999principles}. Quantum estimation theory has shown that this limitation is not fundamental for two equally bright, mutually incoherent sources: the single-parameter quantum Fisher information (QFI) for their separation remains finite even in the zero-separation limit and can be attained using experimentally feasible spatial-mode demultiplexing \cite{tsang2015quantum, tsang2016quantum, 
paur2016O,tsang2019resurgence, tsang2019resolving, 
Lupo2020, Boucher2020, hradil2021exploring, wadood2021experimental, liang2021coherence, 
Zanforlin2022,sajia2022superresolution, sajia2025enhanced, Santamaria2023SpatialMode, thachil2023achieving, darji2024robust, sorelli2024multimode, Rouviere2024UltraSensitive, Lvovsky2026PassiveReview}. Beyond two-point-source estimation, projections onto Hermite--Gaussian
spatial modes have enabled passive far-field superresolution imaging
of coherent and incoherent extended objects, as well as resolution
enhancement under focused illumination
\cite{Pushkina2021Superresolution,Frank2023Passive,
Duplinskiy2025Tsang}. The idealized assumptions behind this conclusion are consequential. Realistic passive sources may have unequal brightnesses and unknown mutual coherence, including an unknown relative coherence phase \cite{Sorelli2022,Kurdzialek2022}. These nuisance parameters are generally coupled to the source separation in the detection statistics \cite{Grace2020}.

Two major challenges arise in realistic multiparameter superresolution. The first concerns the attainable precision: when the nuisance parameters are unknown, {\em how much information about the separation remains accessible?} This question must be addressed through a multiparameter quantum Fisher information analysis. Previous studies have shown that uncertainty in the source properties can substantially modify the attainable precision, although finite separation information may persist in the nonzero sub-Rayleigh regime \cite{Rehacek2017PRA, larson2018resurgence, 
Rehacek2018Optimal, hervas2024optimizing}. The second, and far more pressing (and less explored), challenge is operational: {\em how can this information be extracted without prior knowledge of the nuisance parameters?} Conventional multiparameter strategies generally require their simultaneous estimation, demanding measurements of source properties that are difficult, and often impractical, to access for realistic passive signals.

\begin{figure*}[ht]
\includegraphics[width=2\columnwidth]
{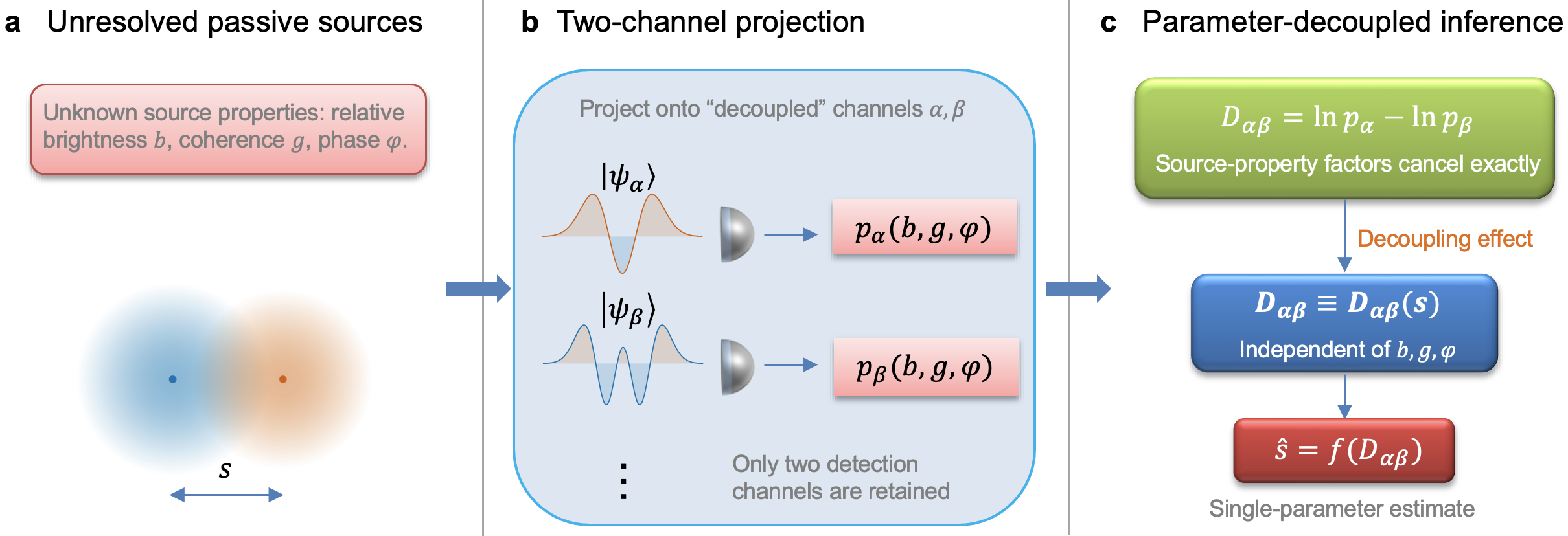}
\caption{
Concept of parameter-decoupled quantum superresolution.
\textbf{(a)} Two passive sources separated by $s$ produce overlapping
point-spread functions. Their brightness imbalance $b$, mutual-coherence
magnitude $g$, and relative phase $\varphi$ are generally unknown. \textbf{(b)} The received field is projected onto two ``decoupled" channels, producing detection probabilities $p_\alpha$ and $p_\beta$ that depends on multiple source parameters $b,g,\varphi$. \textbf{(c)} Their log-probability observable $D_{\alpha\beta}=\ln p_\alpha-\ln p_\beta=D_{\alpha\beta}(s)$ cancels the common dependence on $b$, $g$, and $\varphi$, leaving a function of the separation alone. The separation can therefore be recovered through $\hat{s}=f(D_{\alpha\beta})$ without estimating the three source-property nuisance parameters.}
\label{Fig_scheme}
\end{figure*}

In this article, we address both challenges for a general partially coherent unbalanced two-source state. First, we perform a complete four-parameter quantum Fisher information matrix (QFIM) analysis and derive the nuisance-adjusted QFI for the separation through the Schur complement \cite{yang2019attaining,suzuki2020quantum}. We show that this information vanishes in the zero-separation limit but remains finite at any nonzero separation in the sub-Rayleigh regime, demonstrating that meaningful separation information remains available.

Second, and more importantly, we construct a two-channel log-probability invariant whose ensemble value exactly decouples all three nuisance parameters. We identify two broad classes of measurement channels (function vectors) satisfying the parameter-decoupling condition: (i) parity eigenmodes and (ii) nonparity channels. For the lowest distinct same-parity Hermite–Gaussian pair, $\mathrm{HG}_{0}$--$\mathrm{HG}_{2}$, the separation information per incident signal carried by the conditional channel label asymptotically saturates the nuisance-adjusted QFI in the sub-Rayleigh limit. Thus, although the physical state belongs to a multiparameter family, an appropriately engineered measurement reduces the  problem to a single unknown parameter and remarkably attains the remaining quantum limit without estimating the nuisance parameters. A schematic illustration of our approach is shown in Fig.~\ref{Fig_scheme}.

\section{Two-source model.} We consider a generic two-center signal associated with a physical coordinate $u$, which represents a general displacement such as position $x$, time $t$, frequency $\omega$, etc. The two center states, separated by a distance $s$, are described by $|h_{\pm}\rangle, \langle u|h_{\pm}\rangle=h_{\pm}(u)=h(u\pm s/2)$,
characterizing a $\pm s/2$ shift of the symmetric signal $h(u)$. No specific functional form is assumed for the two individual center signals, and they are generally nonorthogonal,
$\langle h_-|h_+\rangle=\delta(s)\neq0$. Such a general two-center signal can then be described by the mixed state
\beqa
\rho(s)
&=&
\frac{1}{\mathcal N}
\Big[
p_+|h_+\rangle\langle h_+|
+\sqrt{p_+p_-}\gamma|h_+\rangle\langle h_-|
\notag\\
&&+\sqrt{p_+p_-}\gamma^*|h_-\rangle\langle h_+|
+p_-|h_-\rangle\langle h_-|
\Big],
\label{mixedstate}
\eeqa
where $p_{\pm}$ characterize the relative strengths of the two signals and are not individually normalized, i.e., $p_{+}+ p_{-} \neq 1$. Without loss of generality, we take $p_-\geq p_+$ and quantify the brightness imbalance by $b=p_+/p_-\in[0,1]$, where $b=1$ and $b=0$ correspond to balanced and completely imbalanced signals, respectively. The complex parameter $\gamma=ge^{i\varphi}, 0\le g \leq1$, characterizes their mutual coherence, with $\varphi$ the relative phase. The normalization factor is $\mathcal{N} =p_++p_-+
2g\delta(s)\sqrt{p_+p_-}\cos\varphi$. 

The model therefore encompasses arbitrary partial coherence, relative phase, brightness imbalance, and source statistics without imposing a particular form of $h(u)$, and it depends on the four-dimensional parameter vector
\begin{equation}
\bm{\theta}=(s,b,g,\varphi).
\end{equation}

\section{Nuisance-adjusted quantum limit.} The symmetric logarithmic derivatives $L_{\mu}$ satisfy $\partial_{\mu}\rho= \left(L_{\mu}\rho+\rho L_{\mu}
\right)/2$,
and the QFIM for the four parameters $\mu,\nu=s,b,g,\varphi$ is
\begin{equation}
F_{\mu\nu}^{Q}
=
\frac{1}{2}
\Tr\left[
\rho
\left(
L_{\mu}L_{\nu}+L_{\nu}L_{\mu}
\right)
\right].
\label{eq:qfim}
\end{equation}
Partitioning the matrix into the target parameter $s$ for superresolution and nuisance parameters $\bm{\eta}=(b,g,\varphi)$ gives the QFIM as
\begin{equation}
\bm{F}^{Q}
=
\begin{pmatrix}
F_{ss} & \bm{F}_{s\eta}\\
\bm{F}_{\eta s} & \bm{F}_{\eta\eta}
\end{pmatrix},
\end{equation}
where $\bm{F}_{s\eta}=(\bm{F}_{\eta s})^{T}=(F_{sb},F_{sg},F_{s\varphi})$, and $\bm{F}_{\eta\eta}$ is the $3\times3$ matrix for the nuisance parameters. The identifiable nuisance-adjusted QFI for $s$ can be obtained via the generalized Schur complement \cite{carollo2019quantumness, liu2020quantum,yang2019attaining, suzuki2020quantum}
\begin{equation}
F_{s|\bm{\eta}}^{Q}
=
F_{ss}
-
\bm{F}_{s\eta}
(\bm{F}_{\eta\eta})^{+}
\bm{F}_{\eta s},
\label{eq:schur}
\end{equation}
where ``$+$" denotes the Moore-Penrose pseudoinverse \cite{Penrose_1955} when the nuisance block is rank deficient, otherwise represents the conventional inverse operation.

To expose its physical content, we define the orthonormal parity modes
\begin{equation}
\ket{e}
=
\frac{\ket{h_{+}}+\ket{h_{-}}}
{\sqrt{2(1+\delta)}},
\qquad
\ket{o}
=
\frac{\ket{h_{+}}-\ket{h_{-}}}
{\sqrt{2(1-\delta)}}.
\label{eq:parity}
\end{equation}
Although \(\rho\) has rank at most two at any fixed \(s\), its support
moves with \(s\). The separation derivative consequently decomposes
into a component tangent to the two-dimensional support and a component
normal to it. The corresponding single-parameter QFI for s with the nuisance parameters held fixed is then given as
\begin{equation}
F_{ss}
=
F_{ss}^{(\inte)}
+
F_{ss}^{(\ext)},
\label{eq:decomposition}
\end{equation}
where $F_{ss}^{(\inte)}$ and 
$F_{ss}^{(\ext)}$ represent the tangent and normal parts respectively.

For generic identifiable parameters \(0<b<1\) and \(0<g<1\), variations
of \((b,g,\varphi)\) span the complete three-dimensional tangent space
of a mixed qubit within the support. The Schur projection therefore
removes exactly the internal component \cite{supplemental}
\begin{equation}
\bm{F}_{s\eta}
(\bm{F}_{\eta\eta})^{+}
\bm{F}_{\eta s}
=
F_{ss}^{(\inte)}.
\label{eq:keyidentity}
\end{equation}
The nuisance-adjusted QFI can be explicitly achieved as
\begin{equation}
F_{s|\bm{\eta}}^{Q}
=F_{ss}^{(\ext)}.
\end{equation}

For the example of a Gaussian point-spread function (PSF), $h(u)
= (2\pi\sigma^{2})^{-1/4}
\exp\left(-u^{2}/4\sigma^{2}\right)$, one has the overlap $\delta(s)=\exp\left(-s^{2}/8\sigma^{2}\right)$, where $\sigma$ is the width of the PSF. As a result, $F_{s|\bm{\eta}}^{Q}$ can be explicitly expressed as 
\begin{equation}
F_{s|\bm{\eta}}^{Q}
=
\frac{
(1-e^{-4Q})(A-Be^{-2Q})+4Qe^{-2Q}(B-Ae^{-2Q})}{
4\sigma^{2}(1-e^{-4Q})(A+Be^{-2Q})},
\label{NuisanceQFI-PSF}
\end{equation}
where $A(b)=1+b$, $B(b,g,\varphi)=2\sqrt{b}g\cos\varphi$, and $Q=s^2/16\sigma^2$ \cite{supplemental}. Fig.~\ref{Fig_Fisher} (a) shows the dependence of this
nuisance-adjusted QFI on $s$ as the solid blue curve for the 
representative parameters $b=0.4$, $g=0.6$, 
$\varphi=\pi/8$. Although $F_{s|\bm{\eta}}^{Q}$ vanishes as 
$s\to 0$, it remains finite throughout the sub-Rayleigh regime 
$0<s<\sigma$, indicating that nonzero information about the 
separation remains accessible and can be accumulated through 
repeated measurements.

The incoherent case $g=0$ requires separate treatment because it is a singular boundary of the four-parameter model: when the sources are mutually incoherent, the phase $\varphi$ is undefined and is therefore not an identifiable nuisance parameter. If incoherence is imposed as known prior structure, the conventional result $F_{ss}=1/4\sigma^2$
is recovered \cite{tsang2016quantum, supplemental}.

In the deeply sub-Rayleigh regime, $s\ll \sigma$, one can perform a Taylor expansion and achieve

\begin{equation}
F_{s|\bm{\eta}}^{Q}
=
\frac{s^{2}}{32\sigma^{4}}
-\frac{A-B}{768\sigma^{6}(A+B)}s^{4}
+\mathcal{O}(s^{6}).
\end{equation}
Remarkably, the leading first term is independent of the nuisance parameters $\bm{\eta}=(b,g,\varphi)$. Thus, uncertainty in the source properties persists a quadratic sub-Rayleigh limitation only with respect to the separation $s$ at the level of identifiable quantum information.

\begin{figure*}[ht]
\includegraphics[width=2\columnwidth]
{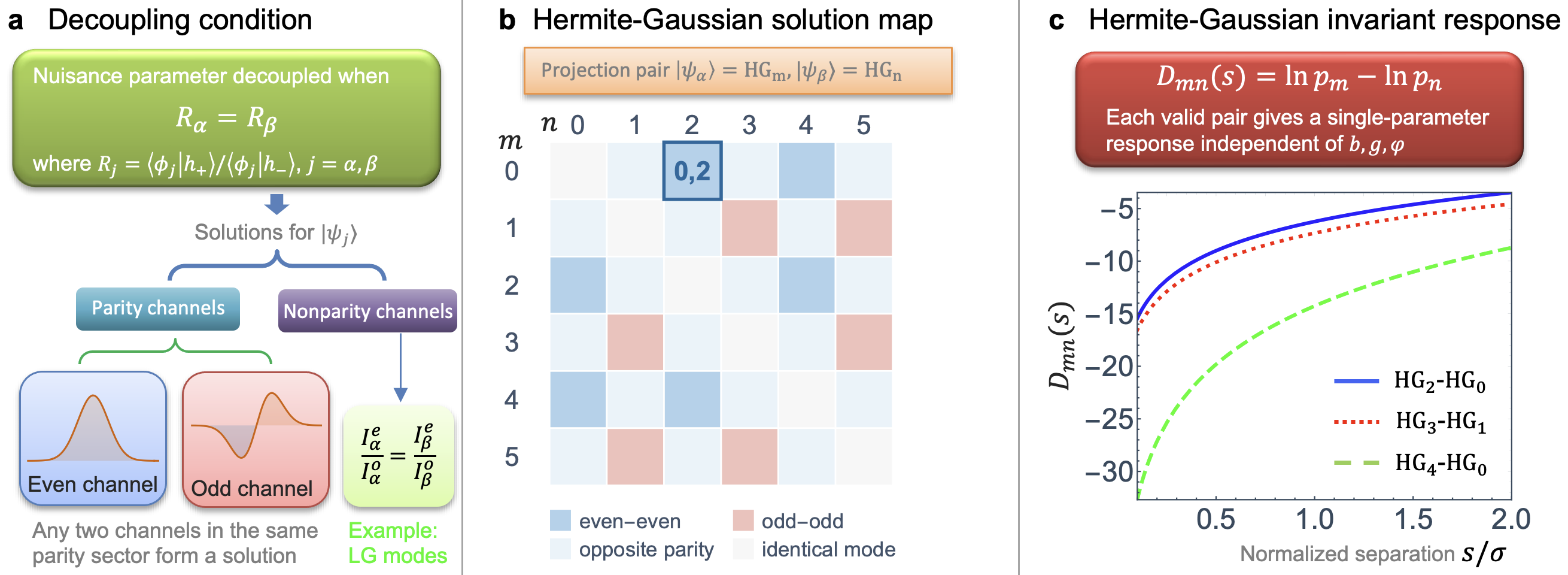}
\caption{General families of parameter-decoupled measurements. \textbf{(a)} For a pair of projection modes $\{\phi_{\alpha},\phi_{\beta}\}$, the dependence on the source-property nuisance parameters cancels when their shifted-PSF overlap ratios satisfy $R_{\alpha}(s)=R_{\beta}(s)$, where $R_j(s)=\langle\phi_j|h_{+}\rangle/\langle\phi_j|h_{-}\rangle$. The general solutions turn into two families : parity eigenmodes and nonparity channels. For parity eigen modes, any two channels belonging to the same parity sector provide a parameter-decoupling solution. \textbf{(b)} Solution map for example pairs of Hermite--Gaussian modes $(\mathrm{HG}_m,\mathrm{HG}_n)$. Even--even and odd--odd pairs satisfy the same-parity condition, whereas opposite-parity pairs do not. The lowest-order nontrivial pair, HG$_0$--HG$_2$, is outlined. \textbf{(c)} Representative separation-dependent probability ratios $D_{mn}(s)=\ln p_m -\ln p_n$ for several valid HG pairs. Each response depends on the separation but is independent of the brightness imbalance $b$, coherence magnitude $g$, and relative phase $\varphi$. }
\label{Fig_decoupling}
\end{figure*}

\section{Parameter-decoupled observable.} Although the nuisance-adjusted QFI remains finite at any nonzero
separation in the sub-Rayleigh regime, the more pressing question is how
this information can be extracted in practice. Simultaneously measuring
and estimating all nuisance parameters is generally impractical for
passive sources. We therefore construct a measurement that accesses the available separation information without the need to estimate the nuisance parameters $b$, $g$, or $\varphi$ \cite{Sorelli2021PRL,Sorelli2021PRA}.

Consider two distinguishable measurement channels
$\ket{\phi_{\alpha}}$, $\ket{\phi_{\beta}}$, with wave functions
$\braket{u|\phi_j} = \phi_j(u)$, where $j=\alpha, \beta$. Their detection
probabilities are $p_j=\braket{\phi_j|\rho|\phi_j}$. We introduce the directly measurable
log-probability contrast
\begin{equation}
D_{\alpha\beta}
\equiv
\ln p_{\alpha}-\ln p_{\beta}.
\label{eq:logcontrast}
\end{equation}
Individually, $p_{\alpha}$ and $p_{\beta}$ generally depend on all four parameters $(s,b,g,\varphi)$. This dependence can nevertheless be eliminated by choosing the measurement channels to satisfy the parameter-decoupling condition 
\begin{equation}
R_{\alpha}(s)=R_{\beta}(s),
\label{decouple_condition}
\end{equation}
where $R_{j}(s)=I^{+}_{j}(s)/I^{-}_{j}(s)$, and $I_{j}^{\pm}(s)= \braket{\phi_{j}|h_{\pm}}=
\int_{-\infty}^{\infty}
\phi_{j}^{*}(u)h(u\pm s/2)du$. A schematic illustration of general and example solutions for parameter decoupling is shown in Fig.~\ref{Fig_decoupling}

Because $|\phi_j\ra$ is determined by the observer, the overlap amplitudes $I_j^{\pm}(s)$ depend only on the
separation $s$. Thus Eq.~\eqref{decouple_condition} causes the common dependence
of $p_{\alpha}$ and $p_{\beta}$ on the brightness imbalance $b$, coherence magnitude $g$, and coherence phase $\varphi$ to cancel exactly in their ratio \cite{supplemental}. Consequently,
\begin{equation}
D_{\alpha\beta}=D_{\alpha\beta}(s)
=
\ln|I_{\alpha}^{-}(s)|^2
-
\ln|I_{\beta}^{-}(s)|^2,
\label{eq:Dinvariant}
\end{equation}
so that \(D_{\alpha\beta}\) is invariant under variations of all three nuisance
parameters. Thus, although the input state \eqref{mixedstate} belongs to
a four-parameter family, the log-probability invariant $D_{\alpha\beta}$ defines an exactly one-parameter statistical model. See example behaviors of $D_{\alpha\beta}$ with respect to $s$ based on Hermite-Gaussian mode projection in Fig.~\ref{Fig_decoupling} (c).

To construct a nuisance-invariant operational statistic, we condition the measurement statistics on events detected in either channel $\alpha$ or channel $\beta$ and discard all other outcomes. Because this postselected  measurement uses only a subset of the outcomes available in the full Hilbert space, it generally retains less information per incident signal than a measurement performed over the complete outcome space. 

The resulting binary conditional probabilities are
\begin{equation}
q_{\alpha}=\frac{p_{\alpha}}{p_{\alpha}+p_{\beta}},
\quad
q_{\beta} =\frac{p_{\beta}}{p_{\alpha}+p_{\beta}}.
\label{eq:conditionalprobabilities}
\end{equation}
with $q_{\alpha}+q_{\beta}=1$. Then the classical Fisher information per conditionally detected event \cite{Fisher1922,Rao1945,Kay1993} is
therefore
\begin{align}
F_{\alpha\beta}^{(\mathrm{cond})}
=\sum_{j=\alpha, \beta}
\frac{\left(\partial_s q_j\right)^2}{q_j}=
q_{\alpha}q_{\beta}
\left(
\frac{\partial D_{\alpha\beta}}{\partial s}
\right)^2.
\label{genericconditionalFI}
\end{align}
It quantifies the separation
information obtainable from the parameter-decoupled binary statistics, without requiring the estimation of any nuisance parameters.

The separation information per incident signal carried by the 
conditional channel label is therefore
\begin{align}
F_{\alpha\beta}^{(\mathrm{inc})}
=
p_{\alpha\beta}
F_{\alpha\beta}^{(\mathrm{cond})}
=
p_{\alpha\beta}q_{\alpha}q_{\beta}
\left(
\frac{\partial D_{\alpha\beta}}{\partial s}
\right)^{2},
\label{eq:incidentFI}
\end{align}
where $p_{\alpha\beta}=p_{\alpha}+p_{\beta}$ is the probability that an incident signal is successfully detected in either channel $\alpha$ or channel $\beta$. Equation~\eqref{eq:incidentFI} 
accounts only for the separation information encoded in the conditional channel label; any additional information carried by the success or failure of the postselection itself is not included.

\section{Parameter-decoupling solutions.} We identify two broad classes of measurement-channel pairs
\(\{\ket{\phi_{\alpha}},\ket{\phi_{\beta}}\}\) that satisfy the decoupling condition
in Eq.~\eqref{decouple_condition}: (i) parity eigenmodes and
(ii) nonparity channels.

For the first class, suppose that both measurement channels are
eigenfunctions of the parity operator \(\hat{\Pi}\):
\begin{equation}
\hat{\Pi}\ket{\phi_j}
=
\lambda_j\ket{\phi_j},
\qquad
\lambda_j=\pm1,
\qquad
j=\alpha, \beta.
\label{eq:parity_eigenfunctions}
\end{equation}
The decoupling condition \eqref{decouple_condition} becomes \cite{supplemental}
\begin{equation}
\lambda_{\alpha}=\lambda_{\beta}.
\label{parity_solution}
\end{equation}
Thus, any two parity eigenmodes belonging to the same parity sector
automatically satisfy the parameter-decoupling condition \eqref{decouple_condition}, see illustration also in Fig.~\ref{Fig_decoupling} (a).

%A natural realization is provided by the Hermite--Gaussian (HG) basis, whose elements obey $\mathrm{HG}_m(-u)=(-1)^m\mathrm{HG}_m(u)$ \cite{born1999principles}. More generally, any two functions within the same parity sector may be written, for example in the even sector, as

A natural realization is provided by the Hermite--Gaussian (HG) basis,
whose elements obey
$\mathrm{HG}_m(-u)=(-1)^m\mathrm{HG}_m(u)$
\cite{born1999principles}. Mode-selective projection and sorting in the HG basis have been implemented in several passive-superresolution
experiments, demonstrating that such measurements are experimentally
accessible in both two-source estimation and extended-object imaging
\cite{Boucher2020,Sorelli2021PRL,Pushkina2021Superresolution, Frank2023Passive}.
More generally, any two functions within the same parity sector may be
written, for example in the even sector, as
\begin{align}
\phi_{\alpha}(u)
=\sum_{k=0}^{\infty}c_{2k}
\mathrm{HG}_{2k}(u),
\phi_{\beta}(u)
=\sum_{k=0}^{\infty}d_{2k}
\mathrm{HG}_{2k}(u),
\end{align}
with $c_{2k}, d_{2k}$ being corresponding coefficients. The simplest implementation consists of two individual HG modes, $\phi_{\alpha}(u)=\mathrm{HG}_m(u),
\phi_{\beta}(u)=\mathrm{HG}_n(u)$,
whose indices satisfy $(-1)^m=(-1)^n$, see illustration in Fig.~\ref{Fig_decoupling} (b).

The second class consists of measurement functions that do not
individually possess definite parity in the coordinate $u$. To characterize these solutions,
we decompose each channel into its even and odd components,
\begin{equation}
\phi_j(u)
=
\phi_j^{e}(u)+\phi_j^{o}(u),
\end{equation}
where $\phi_{\alpha}^{e}(u)=[\phi_{\alpha}(u)+\phi_{\alpha}(-u)]/2, \quad \phi_{\alpha}^{o}(u)=[\phi_{\alpha}(u)-\phi_{\alpha}(-u)]/2$. The general solution to \eqref{decouple_condition} is obtained as \cite{supplemental}
\begin{equation}
I_{\alpha}^e I_{\beta}^o
=
I_{\beta}^e I_{\alpha}^o,
\label{nonparity_solution}
\end{equation}
where $I^e_j=\int_{-\infty}^{+\infty} [\phi^e_j(u)]^{*}h_-(u)du$, etc.
Thus, after selecting one nonparity channel \(\phi_{\alpha}(u)\), a second channel
\(\phi_{\beta}(u)\) may be constructed by choosing its even and odd components to satisfy Eq.~\eqref{nonparity_solution}. This generates a broad family of parameter-decoupled measurements without requiring
either channel to have definite parity. Laguerre--Gaussian (LG) modes provide an important two-dimensional
example of the parameter-decoupled nonparity solutions, see detailed analysis in the Supplemental Materials \cite{supplemental}. A schematic illustration of the nonparity solutions is shown in Fig.~\ref{Fig_decoupling} (a).

\section{Nuisance-free attainment of the quantum limit.} We now demonstrate the operational performance of the parameter-decoupled measurement by evaluating the conditional 
Fisher information in Eqns.~\eqref{genericconditionalFI}-\eqref{eq:incidentFI}. As a concrete example, we consider a two-source signal with a Gaussian
PSF and project it onto two same-parity Hermite-Gaussian modes,
$\ket{\mathrm{HG}_m}$ and $\ket{\mathrm{HG}_n}$ with $(-1)^m=(-1)^n$. The conditional Fisher
information per successfully retained event is \cite{supplemental}
\begin{equation}
F_{mn}^{(\cond)}
=
\frac{4(m-n)^{2}}{s^{2}}
\frac{p_m/p_n}{\left(1+p_m/p_n\right)^{2}}.
\label{eq:Fmn}
\end{equation}
Accounting for the total probability $p_{mn} = p_m+p_n$ that an incident signal is successfully detected in either of the two selected modes, the information per incident signal carried by the conditional mode
label is
\begin{equation}
F_{mn}^{(\mathrm{inc})}
=
\left(p_m+p_n\right)F_{mn}^{(\cond)}.
\label{eq:Fmn_inc}
\end{equation}

\begin{figure*}[ht]
\includegraphics[width=1.5\columnwidth]
{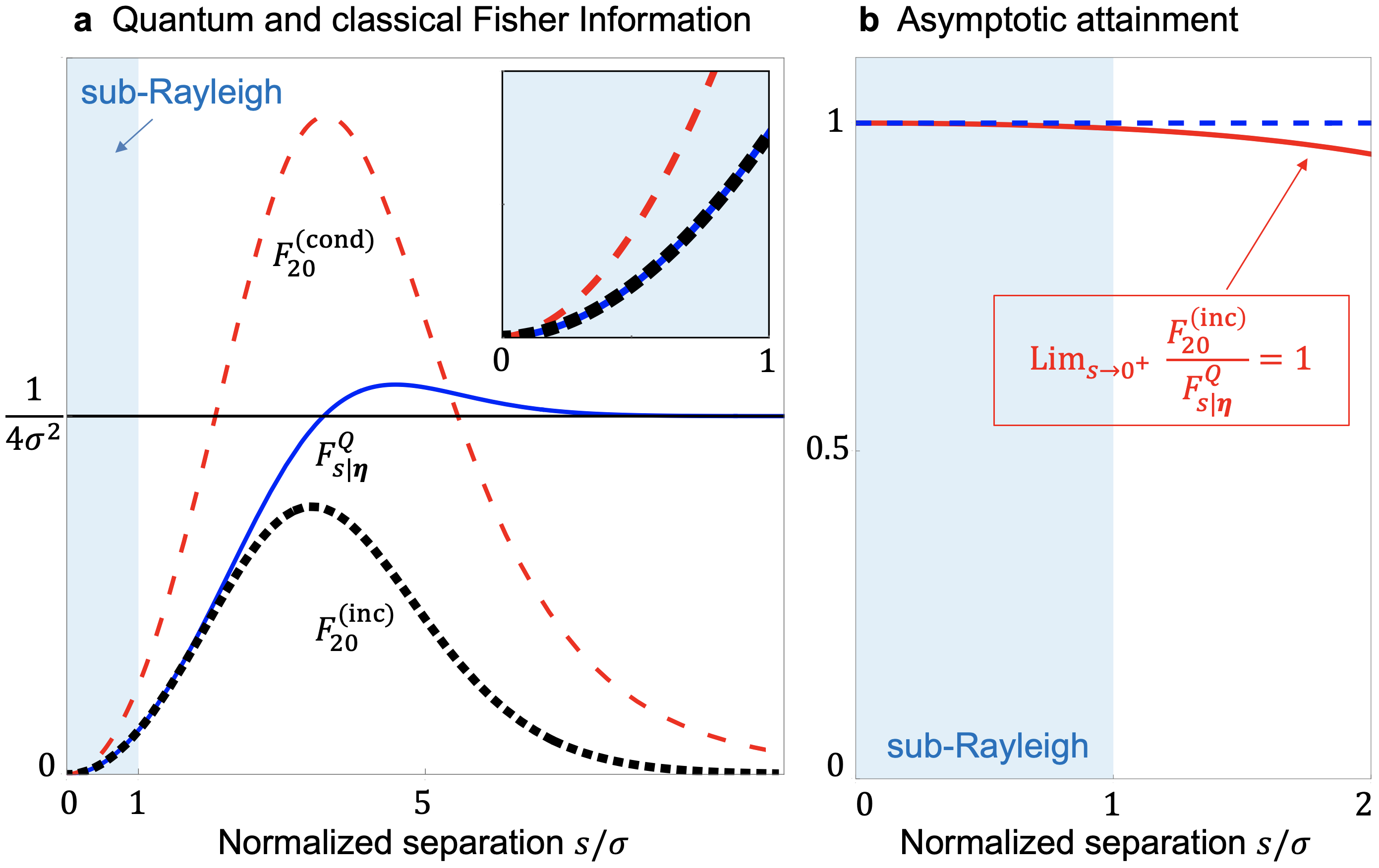}
\caption{Fisher information performance and asymptotic attainment of the quantum limit. \textbf{(a)} Fisher information as a function of normalized separation $s/\sigma$ for the representative source parameters $b=0.4$, $g=0.6$, and $\varphi=\pi/8$. The solid blue curve is the nuisance-adjusted quantum Fisher information $F_{s|\bm{\eta}}^{Q}$ for the full four-parameter model. The red dashed curve is the conditional classical Fisher information $F_{20}^{(\mathrm{cond})}$ per retained HG$_0$-HG$_2$ two-channel event, and the black dotted curve is the corresponding information $F_{20}^{(\mathrm{inc})}$ per incident signal. The shaded region indicates the sub-Rayleigh regime $s<\sigma$. The sub-Rayleigh regime behavior is zoomed in by the upper right panel. \textbf{(b)} Resource-consistent ratio $F_{20}^{(\mathrm{inc})}/F_{s|\bm{\eta}}^{Q}$ in the sub-Rayleigh regime. The ratio approaches unity as $s\rightarrow0^{+}$, demonstrating that the parameter-decoupled measurement asymptotically attains the nuisance-adjusted quantum limit.}
\label{Fig_Fisher}
\end{figure*}

The lowest distinct same-parity pair is $m=2$ and $n=0$. For this pair,
the conditional Fisher information per successful event is given as
\begin{equation}
F_{20}^{(\cond)}
=
\frac{s^{2}}{32\sigma^{4}}
\frac{1}{
\left(
1+s^{4}/512\sigma^{4}
\right)^{2}
},
\label{eq:F20}
\end{equation}
whereas the corresponding information per incident signal is achieved to be 
\begin{equation}
F_{20}^{(\mathrm{inc})}
=
\frac{A+B}{A+Be^{-2Q}}
\frac{e^{-Q}s^2}{
32\sigma^4\left(1+Q^2/2\right)}.
\label{eq:F20_inc}
\end{equation}
See illustration of the behavior of $F_{20}^{(\cond)}$ and $F_{20}^{(\mathrm{inc})}$ with respect to separation $s$ in Fig.~\ref{Fig_Fisher}. 

The conditional Fisher information $F_{20}^{(\cond)}$ is shown by the red dashed curve in Fig.~\ref{Fig_Fisher}. Because it is defined per successfully retained  event, it may exceed the nuisance-adjusted QFI $F_{s|\bm{\eta}}^{Q}$, which is normalized per incident signal. A resource-consistent comparison with the quantum limit must instead use $F_{20}^{(\mathrm{inc})}$, which includes the postselection probability
$p_{20}=p_0+p_2$.

%\caption{Fisher information as a function of the target separation $s$ for $b=0.4$, $g=0.6$, and $\varphi=\pi/8$. The solid blue curve represents the nuisance-adjusted QFI, $F_{s|\bm{\eta}}^{Q}$, for the four-parameter model. The red dashed curve shows the conditional classical Fisher information, $F_{20}^{(\cond)}$,  per successful two-channel detection associated with the parameter-decoupled measurement, while the black dotted curve represents the corresponding Fisher information per incident signal, $F_{20}^{(\mathrm{inc})}$. The enlarged panel in the upper right highlights the asymptotic saturation of $F_{s|\bm{\eta}}^{Q}$ by the operational Fisher information $F_{20}^{(\mathrm{inc})}$ in the sub-Rayleigh regime. The same asymptotic behavior holds for generic interior values of the nuisance parameters $b$, $g$, and $\varphi$. The conditional quantity $F_{20}^{(\cond)}$ may exceed the QFI because it is normalized per retained event; the resource-consistent per-incident quantity $F_{20}^{(\mathrm{inc})}$ remains below the quantum bound and asymptotically approaches it.}

The behavior of $F_{20}^{(\mathrm{inc})}$ is illustrated by the black dotted curve in Fig.~\ref{Fig_Fisher} (a). As expected, it lies below the nuisance-adjusted quantum limit $F_{s|\bm{\eta}}^{Q}$ in the large-separation regime \cite{large-s}. Remarkably, the two quantities converge in the sub-Rayleigh regime. Indeed, for $s\ll\sigma$, the conditional Fisher information has the asymptotic expansion
\begin{equation}
F_{20}^{(\cond)}
=
\frac{s^{2}}{32\sigma^{4}}
+
\mathcal{O}(s^{6}).
\label{eq:F20asymptotic}
\end{equation}
The Fisher information per incident signal has the same leading-order behavior,
\begin{equation}
F_{20}^{(\mathrm{inc})}
=
\frac{s^{2}}{32\sigma^{4}}
-
\frac{A-B}{512\sigma^6(A+B)}s^4
+
\mathcal{O}(s^{6}).
\label{eq:F20_inc_asymptotic}
\end{equation}
Comparing this result with the asymptotic expansion of the nuisance-adjusted QFI yields
\begin{equation}
\lim_{s\rightarrow0^{+}}
\frac{F_{20}^{(\mathrm{inc})}}{F_{s|\bm{\eta}}^{Q}}
=1.
\label{eq:incident_saturation}
\end{equation}
See also its illustration in Fig.~\ref{Fig_Fisher} (b). Thus, the postselection cost does not affect the leading-order separation information in the sub-Rayleigh regime, and the parameter-decoupled $\mathrm{HG}_0$--$\mathrm{HG}_2$ measurement asymptotically saturates the nuisance-adjusted quantum limit on a per-incident-signal basis.

\section{Summary and discussion.} We have introduced a practically realizable parameter-decoupled measurement for extracting the separation information of two realistic passive sources without estimating their unknown brightness imbalance, coherence magnitude, or relative phase. The measurement is constructed from a directly measurable log-probability contrast between two channel-projections. When the channels satisfy a simple parameter-decoupling condition, all nuisance-parameter dependence cancels exactly, reducing the original four-parameter inference problem to an operationally single-parameter measurement. We identified broad classes of decoupling solutions, including parity eigenmodes and nonparity channel pairs.

A complete four-parameter QFIM analysis establishes the corresponding quantum precision limit. The nuisance parameters remove the component of the separation information tangent to the effective two-dimensional support of the state, leaving the normal component as the identifiable nuisance-adjusted QFI. Although this information vanishes at exact coalescence, it remains finite for every nonzero separation in the sub-Rayleigh regime.

For a Gaussian PSF, the lowest-order same-parity pair, $\mathrm{HG}_{0}$-$\mathrm{HG}_{2}$, provides a particularly simple implementation requiring only two channel-resolved detection probabilities. After explicitly accounting for the postselection probability, the resulting Fisher information per incident signal asymptotically saturates the nuisance-adjusted quantum limit in the sub-Rayleigh regime. Thus, the principal result is not merely that identifiable separation information survives in the presence of unknown source properties, but that this information can be extracted through a concrete measurement without estimating those properties at all.

Our framework thereby converts a realistic multiparameter superresolution problem into an experimentally accessible single-parameter inference task. Because the decoupling condition
depends only on shifted-singal overlaps, the approach is not restricted to spatial imaging and may be extended to temporal, spectral, and other two-center resolution problems \cite{De2021,Mazelanik2022,czupryniak2026demonstrating}.The present treatment assumes a calibrated symmetric PSF and a known centroid; extensions to unknown centroids, asymmetric PSFs, and higher-dimensional source configurations provide natural directions for future work.

\noindent{\bf Acknowledgement:} We acknowledge financial support from NSF grants PHY-2316878 and PHY-2514953, and Chin Ying Foundation.

\noindent{\bf Disclosures:} The author declares no conflicts of interest.

\noindent{\bf Data Availability:} No data were generated or analyzed in the presented research.

\bibliographystyle{unsrt}
\bibliography{reference}

\end{document}